# Proton channeling through long chiral carbon nanotubes: the rainbow route to equilibration

S. Petrović*, I. Telečki, D. Borka, and N. Nešković

Laboratory of Physics (010), Vinča Institute of Nuclear Sciences, P. O. Box 522, 11001 Belgrade, Serbia

* Corresponding author, e-mail address: petrovs@vin.bg.ac.yu

**Abstract**

In this work we investigate the rainbows appearing in channeling of 1 GeV protons through the long (11, 9) single-wall carbon nanotubes. The nanotube length is varied from 10 to 500 μm. The angular distributions of channeled protons are computed using the numerical solution of the proton equations of motion in the transverse plane and the Monte Carlo method. The rainbows are identified as the rings in the angular distributions, which correspond to the extrema of the proton deflection functions. Each rainbow is characterized by a sharp decrease of the proton yield on its large angle side. As the nanotube length increases, the number of rainbows increases and the average distance between them decreases in an easily predictable way. When the average distance between the rainbows becomes smaller than the resolution of the angular distribution, one cannot distinguish between the adjacent rainbows, and the angular distribution becomes equilibrated. We call this route to equilibration the rainbow route to equilibration. This work is a demonstration of how a simple one-dimensional bound dynamic system can exhibit a complex collective behavior.

Channeling of high energy positively charged particles through carbon nanotubes was foreseen by Klimov and Letokhov [1]. They showed that a positron beam of the energy of 1 GeV was captured by the nanotube and emitted the X-rays of the energy of about 0.3 MeV in the direction of its propagation. After that, several groups investigated the possibility of guiding high energy positively charged particles with straight and bent nanotubes [2-4]. Greenenko and Shulga [5] were the first to examine in detail the high energy ion motion through straight and bent nanotubes.

Krasheninnikov and Nordlund [6] performed a molecular dynamics study of channeling of low energy heavy ions through nanotubes. Recently, channeling of low, medium and high energy ions through nanotubes was reviewed by Mišković [7].

Petrović et al. applied the theory of crystal rainbows, which had been demonstrated to be the proper theory of ion channeling through thin crystals [8, 9], to channeling of 1 GeV protons through the straight and bent (10, 10) single-wall carbon nanotubes [10, 11]. They showed that the rainbow patterns provide the full explanation of the angular distributions of channeled protons. Recently, in a study of channeling of 0.233 MeV protons through the (11, 9) single-wall carbon nanotubes, Borka et al. demonstrated that the image force acting on the protons gave rise to the additional rainbows in their angular distributions [12].

It must be noted that all the above mentioned studies were theoretical. The direct experimental evidence of ion channeling through carbon nanotubes is still lacking, although some progress along that direction has been reported [13]. Recently, channeling of 300 keV electrons through multi-wall carbon nanotubes was observed experimentally [14].

In this work we investigate theoreticaly the rainbow effect in channeling of 1 GeV protons through the long (11, 9) single-wall carbon nanotubes. The nanotube length is varied from 10 to 500 μm. Our motivation was to find out how does the angular distribution of protons channeled through a chiral nanotube evolve with the nanotube length, and how does it look when the nanotube is very long.

The system we consider is a proton channeled through a (11, 9) single-wall carbon nanotube. The coordinate system is chosen so that the z axis coincides with the nanotube axis and the xy plane coincides with the entrance plane of the nanotube, which is the impact parameter plane. The initial proton velocity is taken to be parallel to the z axis. We assume that the proton motion through the nanotube is determined by the proton-nanotube interaction potential obtained via the Molière approximation of the Thomas-Fermi proton-nanotube atom interaction potential averaged along the axial and azimuthal components of the proton position. The axial averaging means that we use the continuum approximation [15] while the azimuthal averaging is applied since the nanotube is chiral. The average proton-nanotube interaction potential is given by

$$U(\rho) = \frac{16\pi Z_1 Z_2 e^2 R}{3\sqrt{3} d^2} \sum_{i=1}^{3} \alpha_i K_0(\beta_i R/a) I_0(\beta_i \rho/a), \qquad (1)$$

where $Z_1$ and $Z_2$ are the atomic numbers of the proton and nanotube atom, respectively, $e$ is the elementary charge, $R$ is the nanotube radius, $d$ is the nanotube atoms bond length, $\rho = \sqrt{x^2 + y^2}$ is the radial component of the proton position, $x$ and $y$ are the transverse rectangular components of the proton position, $a = [9\pi^2/(128 Z_2)]^{1/3} a_0$ is the nanotube atom screening radius, and $a_0$ is the Bohr radius; $I_0$ and $K_0$ designate the modified Bessel functions of the first and second kinds and zero order, respectively, and $(\alpha_i) = $ (0.35, 0.55, 0.10) and $(\beta_i) = $ (0.30, 1.20, 6.00) are the fitting parameters [16].

The effect of thermal vibrations of the nanotube atoms changes expression (1) into

$$U^{th}(\rho) = U(\rho) + \frac{\sigma_{th}^2}{2}(\partial_{xx} + \partial_{yy}) U(\rho), \qquad (2)$$

where $\sigma_{th}$ is the one-dimensional thermal vibration amplitude of the nanotube atoms [8]. Thus, the average proton-nanotube interaction potential with the effect of thermal vibrations included is given by

$$U^{th}(\rho) = \frac{16\pi Z_1 Z_2 e^2 R}{3\sqrt{3} d^2} \sum_{i=1}^{3} (\alpha_i + \frac{\sigma_{th}^2 \beta_i^2}{2a^2}) K_0(\beta_i R/a) I_0(\beta_i \rho/a). \qquad (3)$$

The angular distribution of channeled protons is obtained using the numerical solution of the proton equations of motion in the transverse plane and the Monte Carlo method. The rectangular components of the proton impact parameter are chosen randomly within the circle around the origin of radius $R - a$ [8]. We do not take into account the uncertainty of the proton scattering angle caused by its collisions with the nanotube electrons and the proton energy loss.

The angular distribution of channeled protons can be analyzed via the mapping of the impact parameter plane, the $xy$ plane, to the scattering angle plane, the $\Theta_x \Theta_y$ plane [8]. Since the proton scattering angle is small (smaller then the critical angle for channeling), the proton differential transmission cross section is given by

$$\sigma = \frac{1}{|J|}, \qquad (4)$$

where $J = \partial_x \Theta_x \partial_y \Theta_y - \partial_x \Theta_y \partial_y \Theta_x$ is the Jacobian of the mapping [17, 18]. Therefore, equation $J = 0$ determines the lines in the impact parameter along which the proton differential transmission cross section diverges. These lines are called the rainbow lines in the impact parameter plane. The rainbow lines in the scattering angle plane are the images of the rainbow lines in the impact parameter plane defined by the mapping.

Due to the cylindrical symmetry of the average proton-nanotube interaction potential, the problem we consider is in fact one-dimensional, and the rainbow lines in the impact parameter plane and scattering angle plane will show up as the circles. Consequently, it is sufficient to analyze the mapping of the $x$ axis in the impact parameter plane to the $\Theta_x$ axis in the scattering angle plane, i.e., the $\Theta_x(x)$ deflection function. The abscissas and ordinates of the extremal points of the deflection function determine the radii of the rainbow lines in the impact parameter plane and scattering angle plane, respectively.

Figure 1 shows the average proton-nanotube interaction potential with the effect of thermal vibrations of the nanotube atoms included in an (11, 9) single-wall carbon nanotube as a function of the radial transverse component of the proton position, which is varied between $-(R-a)$ and $R-a$. The nanotube atoms bond length is 0.14 nm and, hence, the nanotube radius is 0.69 nm [19]. The nanotube atom screening radius is 0.026 nm. The one-dimensional thermal vibration amplitude, estimated using the Debye approximation, is 0.0053 nm [20]. It is obvious that the average interaction potential has the form of a well. Let us mention that it does not have any inflection point within the region of variation of the radial transverse component of the proton position [21]. Since the initial radial component of the proton velocity is equal to zero, the transverse motion of each proton is the one-dimensional oscillatory motion along the line passing through the nanotube center.

Figures 2(a)-(d) show the angular distributions of channeled protons along the $\Theta_x$ axis for the nanotube lengths of 10, 50, 100 and 500 μm, respectively. The size of a bin along the $\Theta_x$ axis is 0.866 μrad and the initial number of protons is 16,656,140. These values enabled us to attain a high resolution of the angular distributions in a reasonable computational time. The angular distributions obtained for the nanotube lengths of 10, 50 and 100 μm contain a central maximum and a number of symmetric pairs of maxima characterized by a sharp decrease of the proton yield on the large angle side; the numbers of symmetric pairs of sharp maxima are one, eight and 15, respectively. The angular distribution obtained for the nanotube length of 500 μm contains a central maximum and a large number of symmetric pairs of sharp maxima close to each other. The analysis shows that in the region where $|\Theta_x| \leq 0.2$ mrad one can identify 55 such pairs of maxima. Figure 2(d) also shows that in the region where 0.041 mrad $\leq |\Theta_x| \leq$ 0.058 mrad there are five pairs of maxima; they are designated by 11-15. In the region in which $|\Theta_x| > 0.2$ mrad the resolution of the angular distribution is not sufficiently high to distinguish easily between the adjacent pairs of maxima.

Figure 3 gives the angular distribution of channeled protons for the nanotube length of 50 μm. It contains a central maximum and eight circular maxima characterized by a sharp decrease of the proton yield on the large angle side, which correspond to a central maximum and the eight symmetric pairs of sharp maxima given in Fig. 2(b). The sizes of a bin along the $\Theta_x$ and $\Theta_y$ axes are 5.6 μrad.

Figures 4(a)-(d) give the mappings of the $x$ axis in the impact parameter plane to the $\Theta_x$ axis in the scattering angle plane, i.e., the $\Theta_x(x)$ deflection functions, for the nanotube lengths of 10, 50, 100 and 500 μm, respectively. The deflection functions obtained for the nanotube lengths of 10, 50 and 100 μm contain one, eight and 15 symmetric pairs of extrema, respectively; each pair contains a minimum and maximum. The analysis shows that the deflection function obtained for the nanotube length of 500 μm contains 76 symmetric pairs of such extrema. Figure 4(d) also gives five such pairs of extrema, designated by 11-15, that correspond to the five maxima shown in Fig. 2(d). Each symmetric pair of extrema defines a circular rainbow line in the impact parameter plane and a circular rainbow line in the scattering angle plane. It should be noted that each of the deflection functions obtained for the nanotube lengths

of 50, 100 and 500 μm has two envelopes, one of them connecting its extrema designated by odd numbers and the other connecting its extrema designated by even numbers. One can say that, in fact, each of these deflection functions oscillates between the two envelopes.

We also calculated the angular distributions of channeled protons along the $\Theta_x$ axis for the nanotube lengths of 10, 50, 100 and 500 μm and the accompanying $\Theta_x(x)$ deflection functions without the effect of thermal vibrations of the nanotube atoms included. They differ from the corresponding dependences with the effect of thermal vibrations included very slightly. This is explained by the facts that the pairs of extrema of each deflection function corresponding to the dominant pairs of maxima of the accompanying angular distribution are positioned far from the nanotube wall, which is defined by the nanotube atom screening radius [see Figs. 4(a)-(d)], and that the one-dimensional thermal vibration amplitude is much smaller than the screening radius (see Fig. 1).

The comparison of Figs. 2(a)-(c) and Figs. 4(a)-(c) clearly shows that the abscissae of the pairs of maxima of each angular distribution of channeled protons along the $\Theta_x$ axis coincide with the ordinates of the corresponding pairs of extrema of the accompanying $\Theta_x(x)$ deflection function. One can conclude the same for the pairs of maxima of the angular distribution given in Fig. 2(d) and the corresponding pairs of extrema of the deflection function given in Fig. 4(d), in spite of the fact that in the region where $|\Theta_x| > 0.2$ mrad it is not possible to identify the additional 21 pairs of maxima of the angular distribution. Hence, we conclude that the maxima of the angular distributions can be attributed to the rainbow effect. In fact, they are the rainbow singularities. The small angle side of each of these rainbows is its bright side while its large angle side is its dark side.

The analysis of the proton trajectories corresponding to the pair of extrema of each proton deflection function designated by $m$, i.e., the $m^{th}$ pair of rainbow trajectories, shows that they comprise $n_m$ deflections from the nanotube wall [11]. This means that the corresponding rainbow is the rainbow of the $n_m^{th}$ order. The relation between these two integers is $n_m = n_1 + m - 1$, where $n_1$ is the number of deflections from the nanotube wall comprised by the first pair of rainbow trajectories ($m = 1$). It is interesting to note that for the nanotube lengths of 10, 50, 100 and 500 μm $n_1$ = 1, 2, 3 and 13, respectively, meaning that the corresponding rainbows are the rainbows of the first, second, third and 13$^{th}$ order, respectively.

Let us now examine the facts that the number of maxima of the angular distribution of channeled protons along the $\Theta_x$ axis in the region $\Theta_x \geq 0$ or $\Theta_x \leq 0$ with the central maximum excluded, i.e., the number of rainbows, increases and the average distance between the neighboring rainbows decreases as the nanotube length increases. These facts are seen clearly in Figs. 2(a)-(d).

Figure 5 shows the dependences of the number of rainbows and the average distance between them on the nanotube length. In order to avoid the problem of poor resolution of the angular distributions in the region where $\Theta_x$ is large, the numbers and positions of the rainbows are determined from the accompanying $\Theta_x(x)$ deflection functions. The analysis shows that the former dependence can be reproduced excellently by fitting function $f_1 = a_1 L$, where $L$ is the nanotube length and $a_1$ = 0.15 is the fitting parameter. The latter dependence can be reproduced excellently by fitting function $f_2 = a_2 \exp(-L/b_2) + a_3 \exp(-L/b_3)$, where $a_2$ = 0.072 mrad, $b_2$ = 42.6 μm, $a_3$ = 0.015 mrad and $b_3$ = 340.5 μm are the fitting parameters. These fitting functions are also shown in Fig. 5. When the nanotube becomes sufficiently long for the average distance between the rainbows to become smaller than the resolution of the angular distribution, one cannot distinguish between the adjacent rainbows. This means that the rainbows disappear and the angular distribution becomes a bell-shaped one. The analysis shows that, in parallel, the spatial distribution of channeled protons in the exit plane of the nanotube also becomes a bell-shaped one. Hence, one can say that, when the nanotube becomes sufficiently long, the angular and spatial distributions equilibrate, and, as one would expect, this does not happen in accordance with the ergodic hypothesis [22]. We call this route to equilibration, which is characterized by the linear increase of the number of rainbows and the exponential decrease of the distance between them with the increase of the nanotube length, the rainbow route to equilibration.

In this work channeling of 1 GeV protons through the long (11, 9) single-wall carbon nanotubes and the accompanying rainbows are investigated. The nanotube length is varied from 10 to 500 μm. An interesting behavior of the angular distributions of channeled protons is observed – they contain the rings whose number increases and the average distance between them decreases as the nanotube length increases. Each ring is characterized by a sharp decrease of the proton yield on its large angle side, and it is recognized as a rainbow. These rainbows appear in spite of the fact that the average proton-nanotube interaction potential does not have any inflection point within the region of variation of the radial transverse component of the proton position [21]. When the average distance between the rainbows becomes smaller than the resolution of the angular distribution, it becomes equilibrated, in parallel with the equilibration of the spatial distribution of channeled protons in the exit plane of the nanotube. We call this route to equilibration the rainbow route to equilibration. It does not include the uncertainty of the proton scattering angle caused by its collisions with the nanotube electrons and the proton energy loss. This result is an excellent example of a complex collective behavior of a simple one-dimensional bound dynamic system.

We believe that the angular distributions of channeled protons for other types of chiral carbon nanotubes become equilibrated in a similar way as the angular distribution for the type of nanotube considered here, since in all these cases the proton motion is determined by the same type of average proton-nanotube interaction potential. In addition, one may hypothesize that the rainbow route to equilibration is a general characteristic of one-dimensional bound dynamic systems when the time tends to infinity.

# Figure captions

Figure 1. The average proton-nanotube interaction potential with the effect of thermal vibrations of the nanotube atoms included in an (11, 9) single-wall carbon nanotube as a function of the radial transverse component of the proton position. $R$ is the nanotube radius and $a$ the nanotube atom screening radius.

Figure 2. The angular distributions along the $\Theta_x$ axis of 1 GeV protons channeled through the (11, 9) single-wall carbon nanotubes of the length of (a) 10 μm, (b) 50 μm, (c) 100 μm and (d) 500 μm.

Figure 3. The angular distribution of 1 GeV protons channeled through the (11, 9) single-wall carbon nanotube of the length of 50 μm.

Figure 4. The mappings of the $x$ axis in the impact parameter plane to the $\Theta_x$ axis in the scattering angle plane for 1 GeV protons channeled through the (11, 9) single-wall carbon nanotubes of the length of (a) 10 μm, (b) 50 μm, (c) 100 μm and (d) 500 μm.

Figure 5. The dependences of the number of rainbows appearing in the angular distributions of 1 GeV protons channeled through the (11, 9) single-wall carbon nanotubes – $N$, open circles, and the average distance between them – $\delta$, closed circles, on the nanotube length – $L$. The full lines represent the fitting curves.

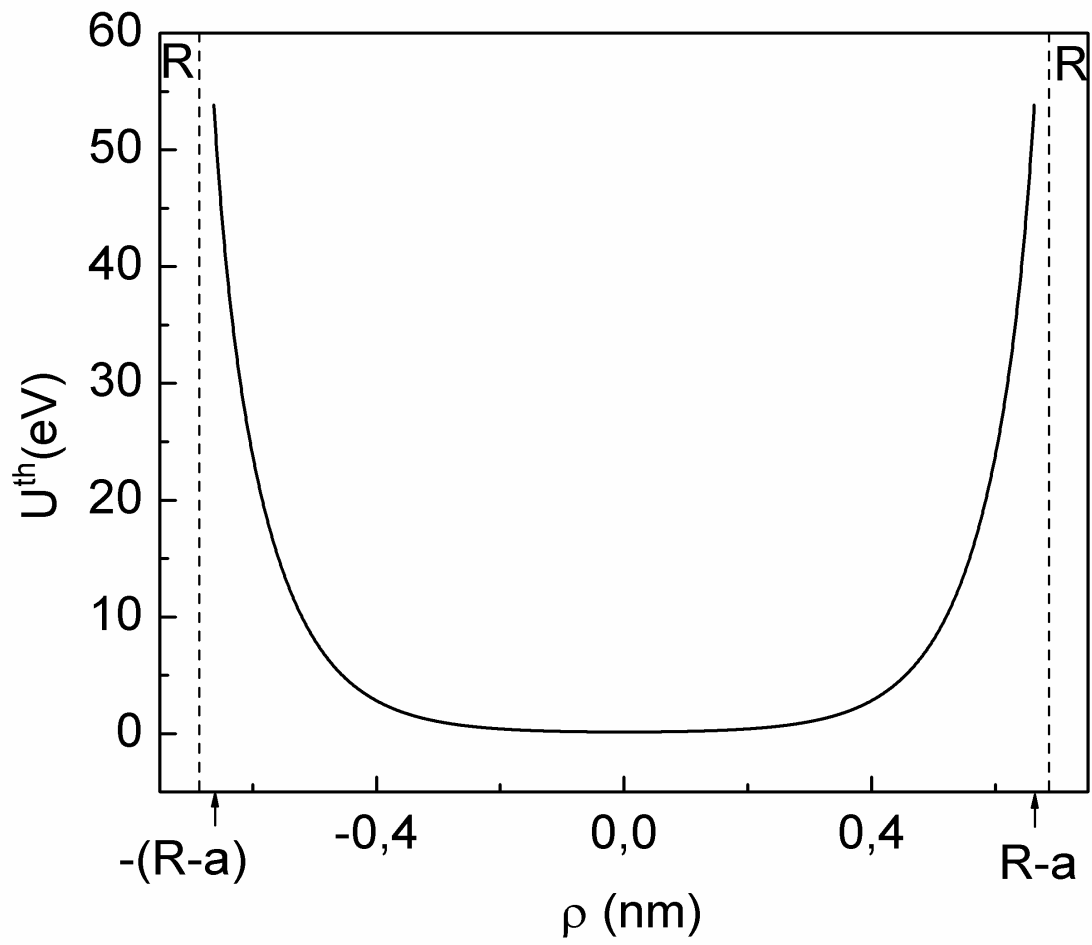

Figure 1

Figure 2(a)

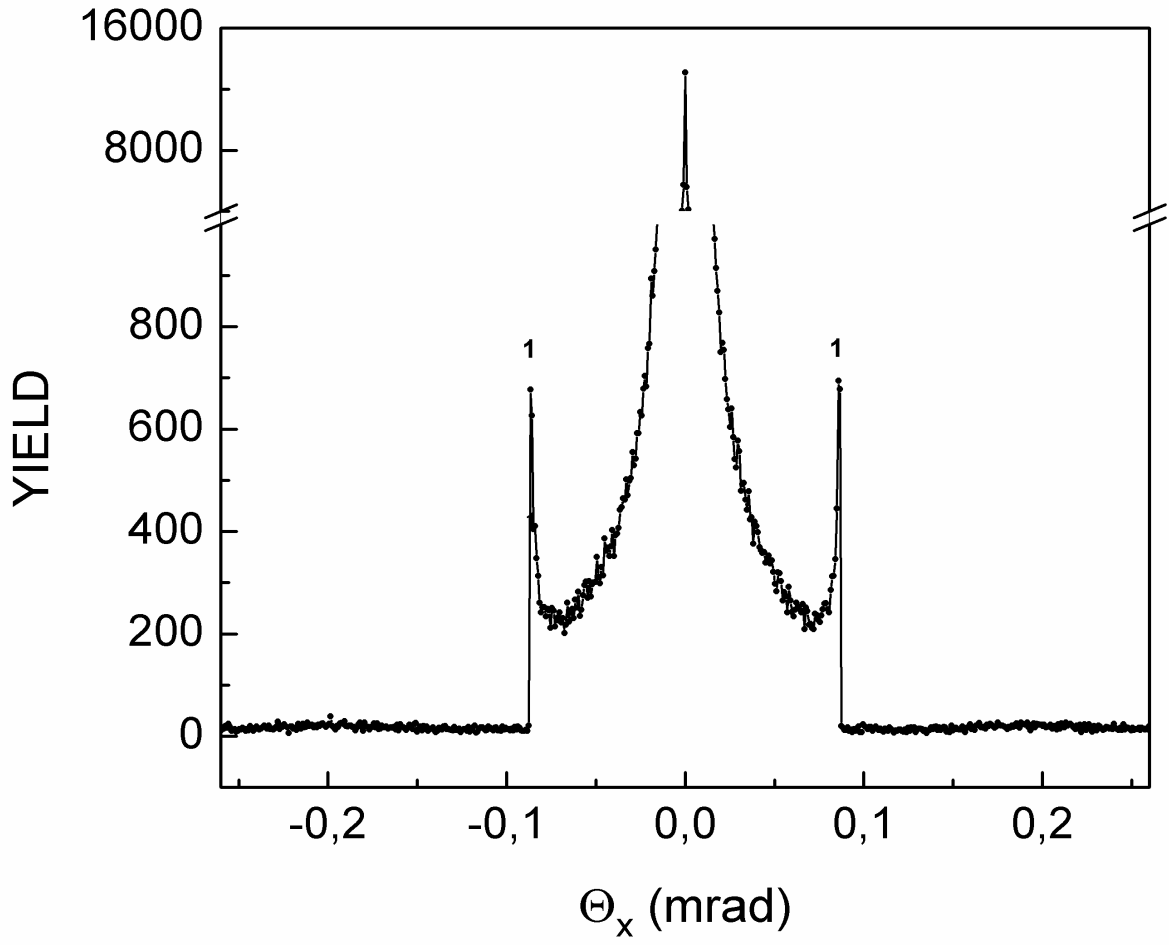

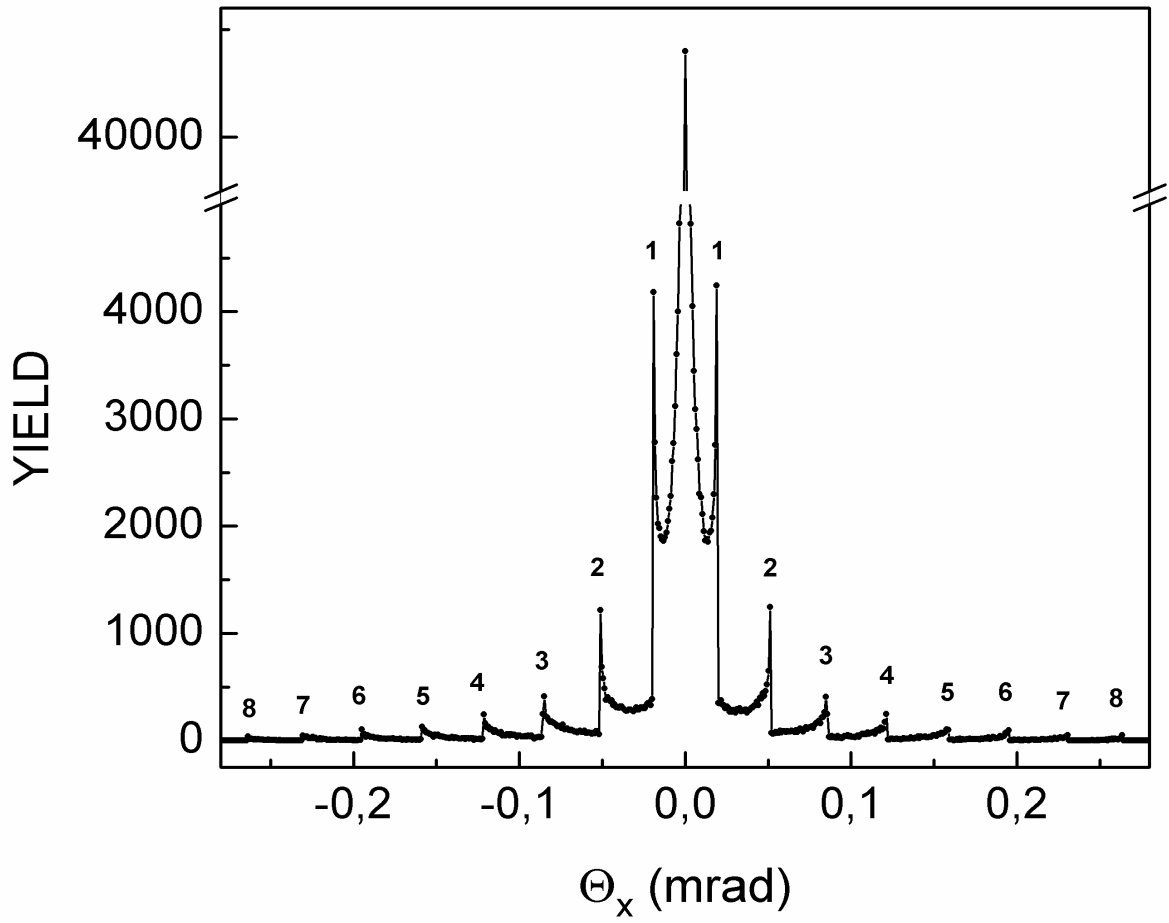

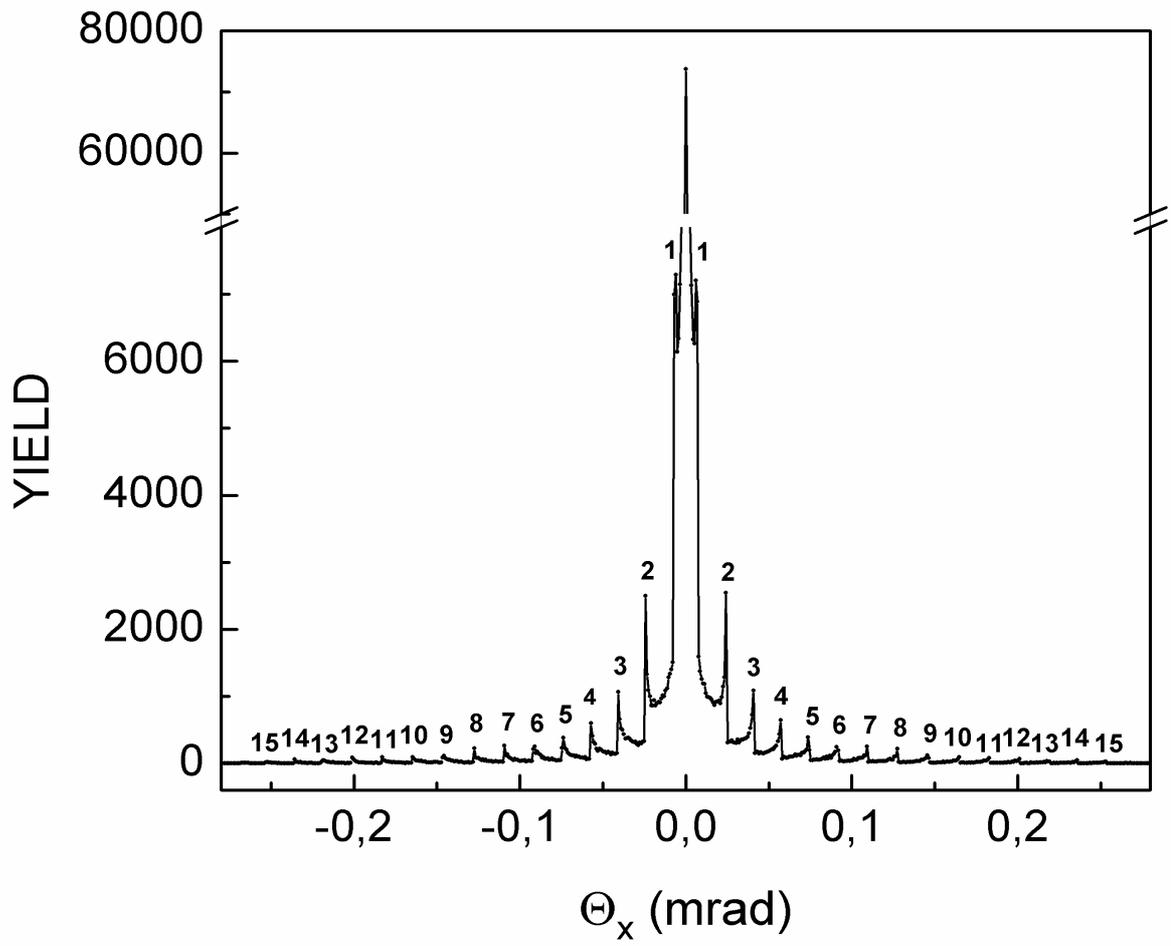

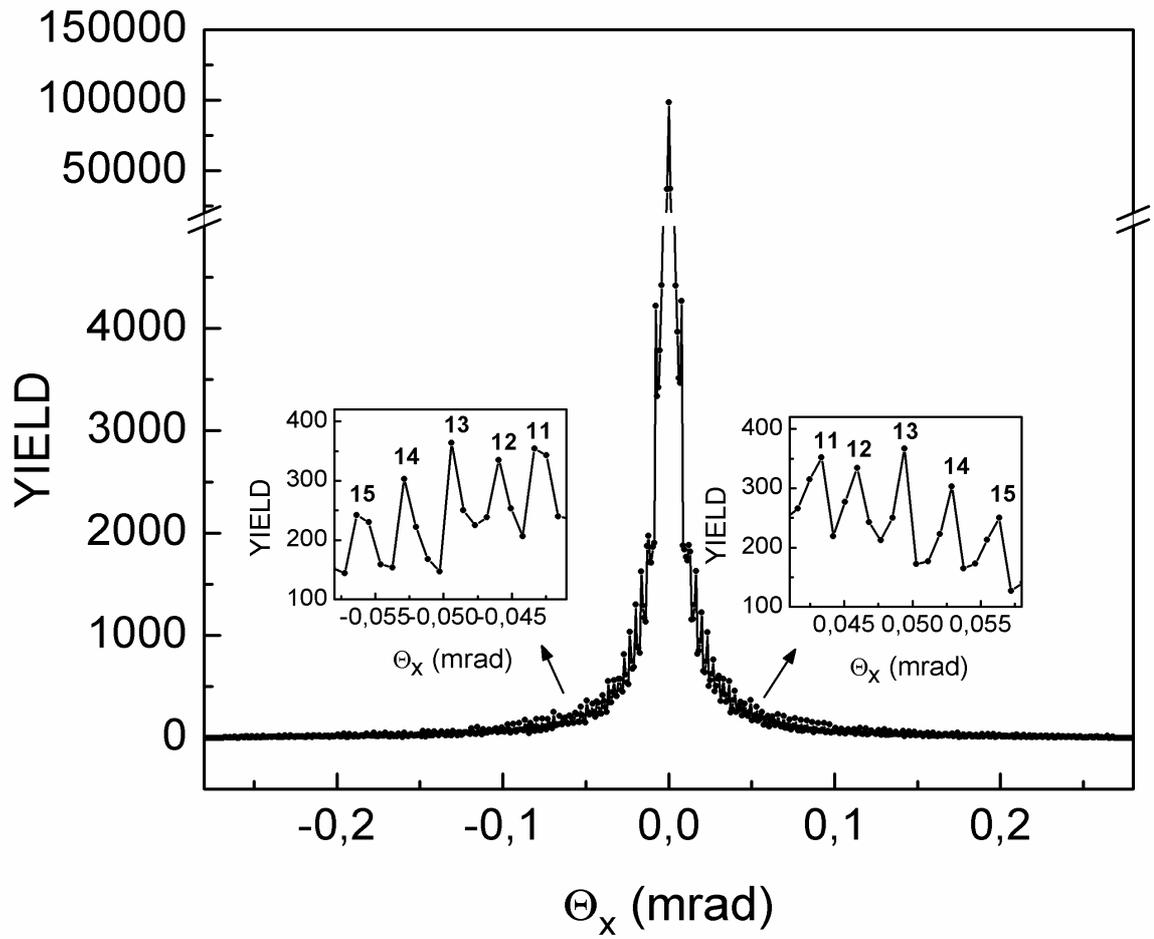

Figure 2(d)

Figure 3

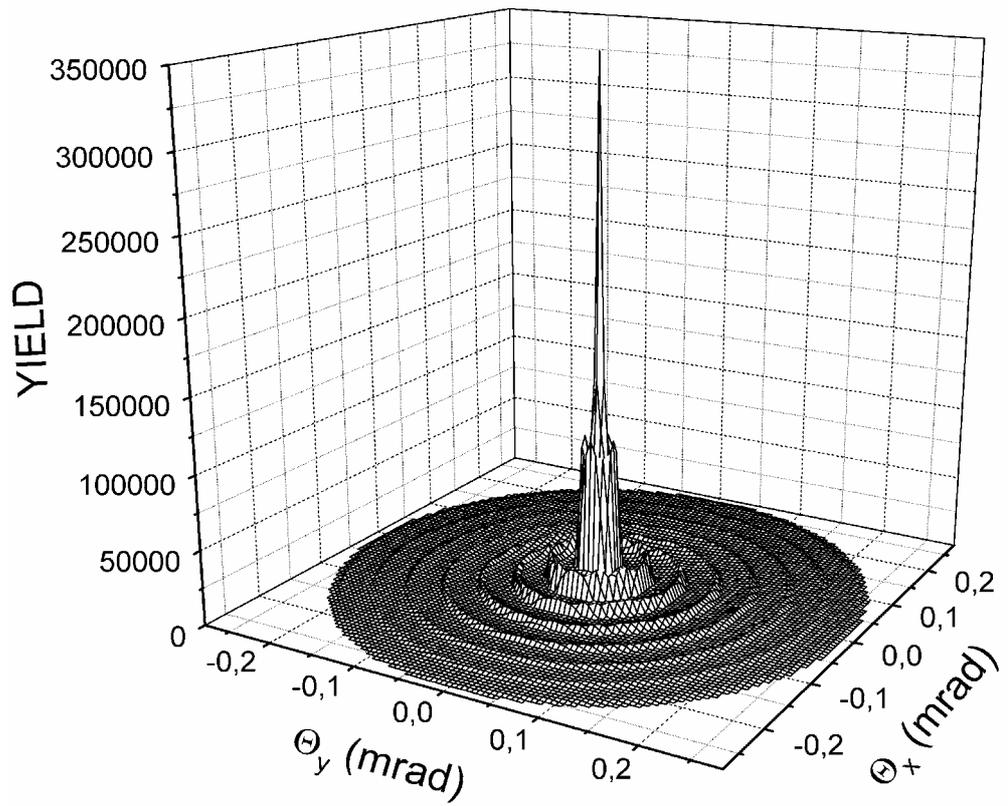

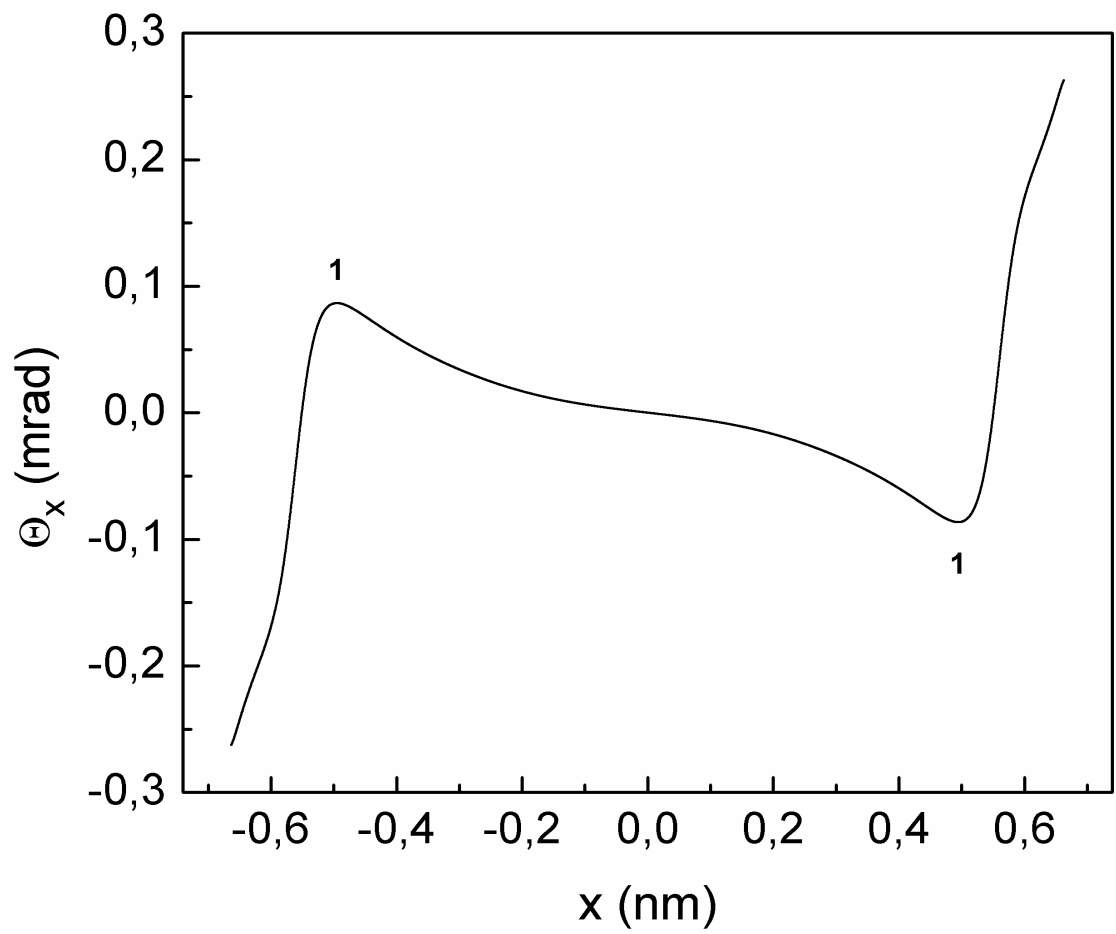

Figure 4(a)

Figure 4(b)

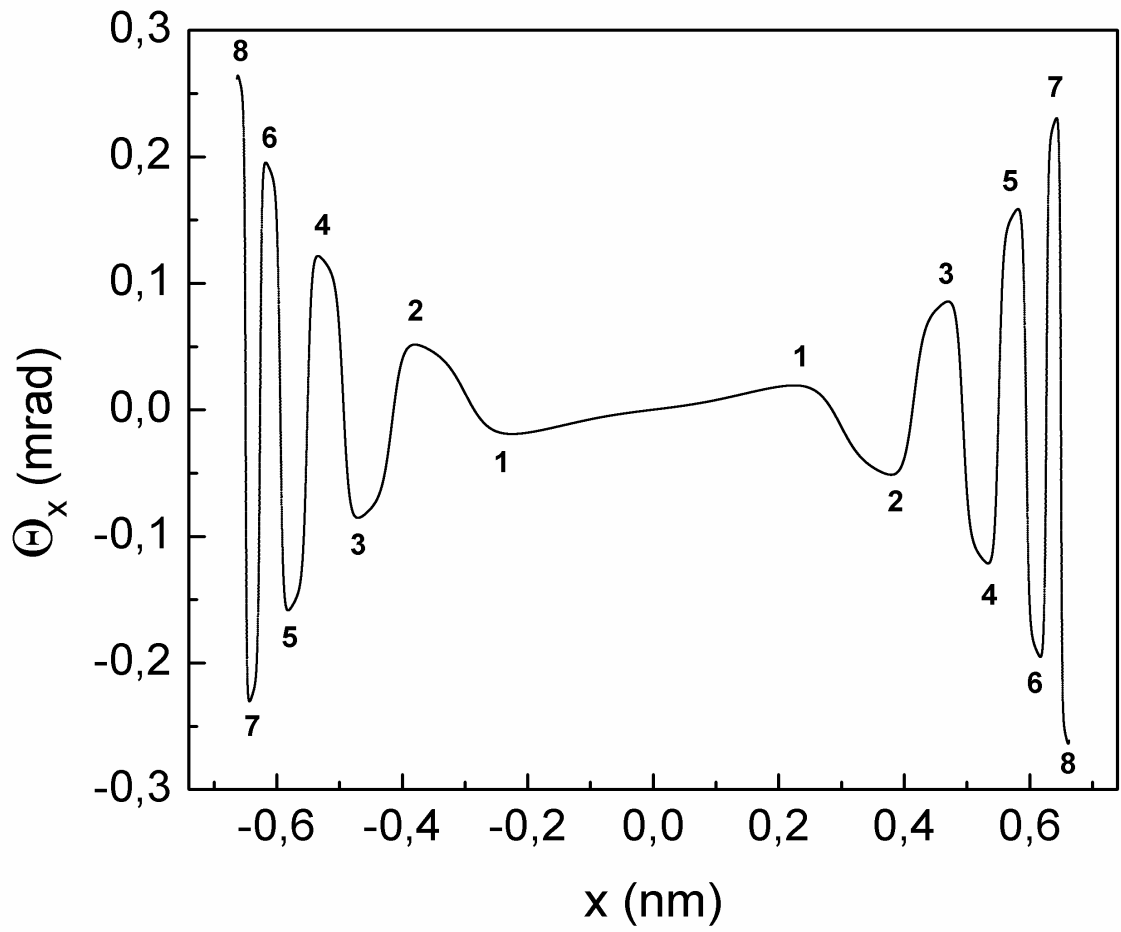

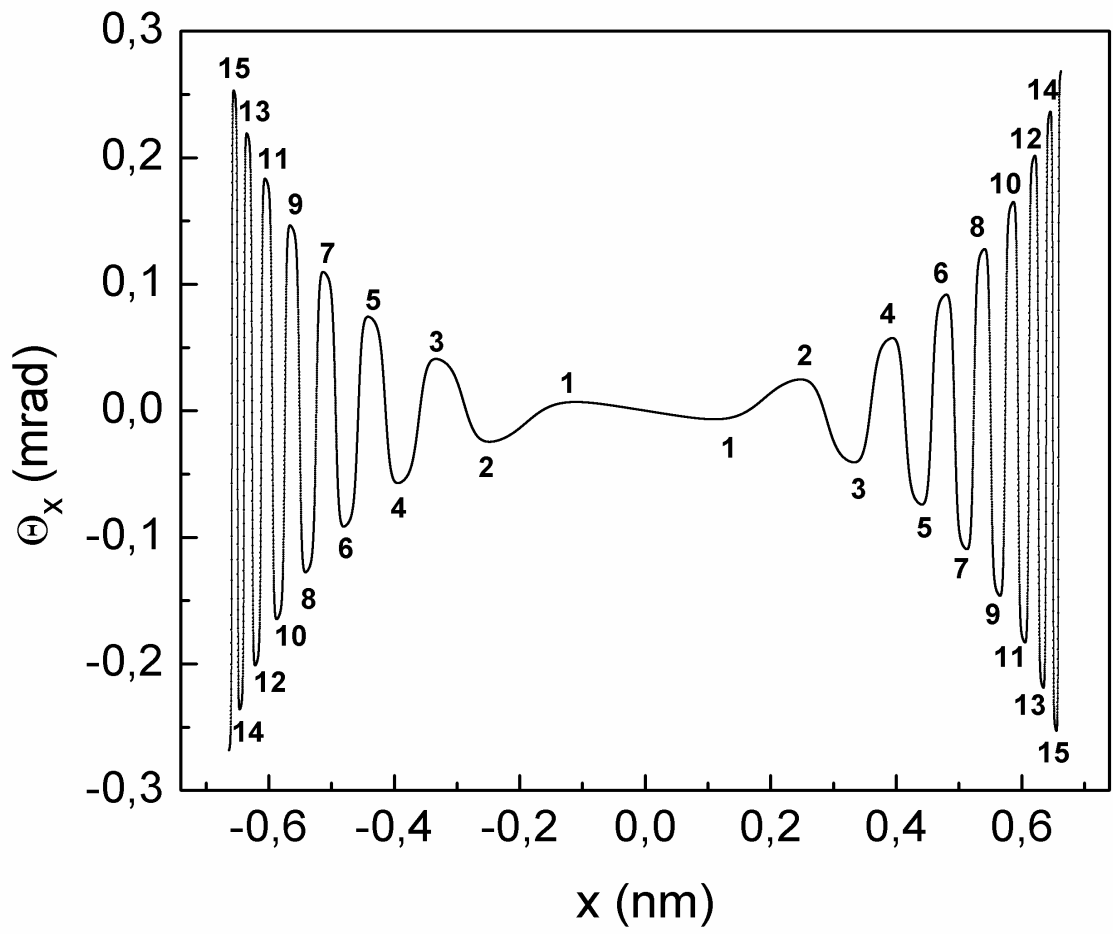

Figure 4(c)

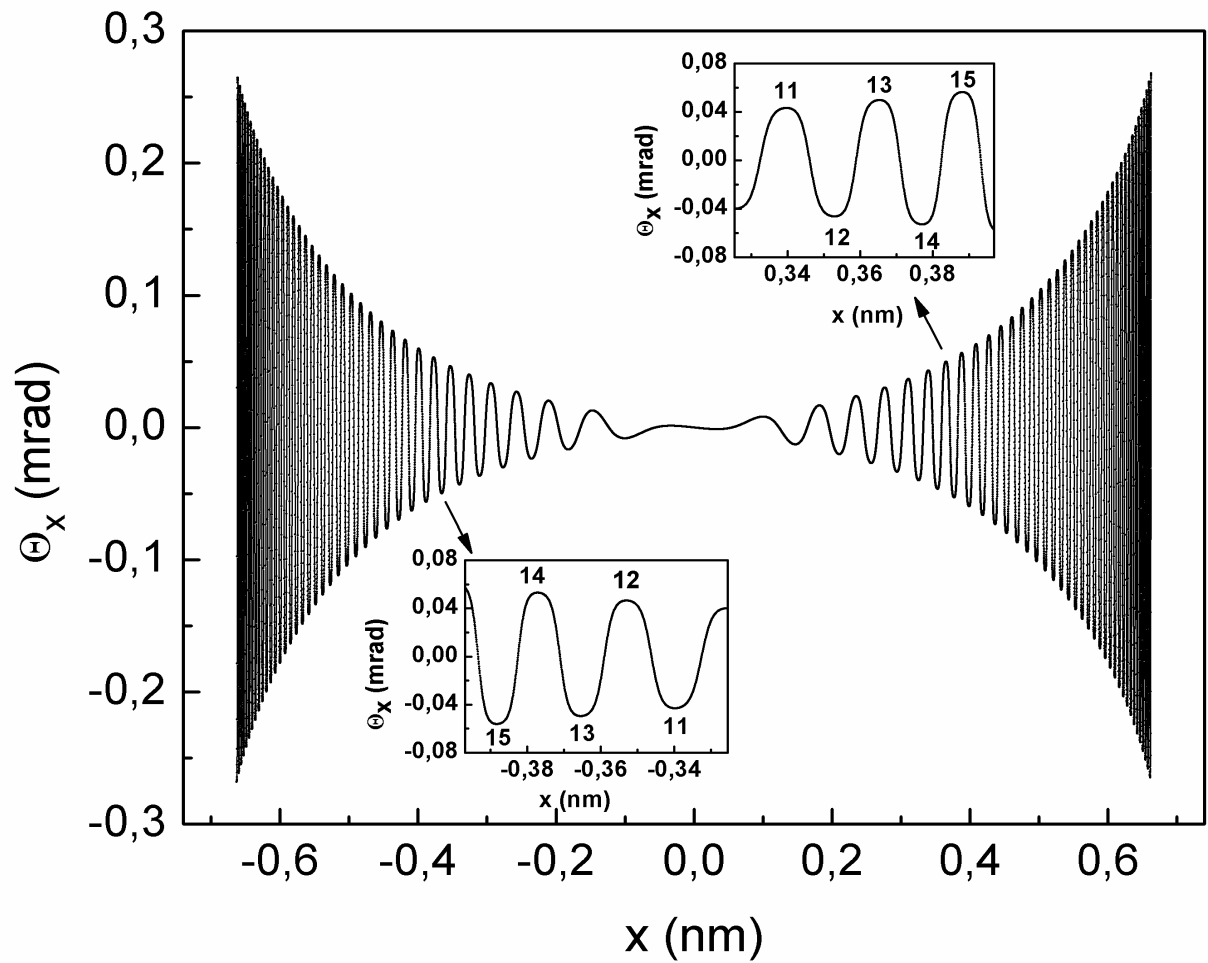

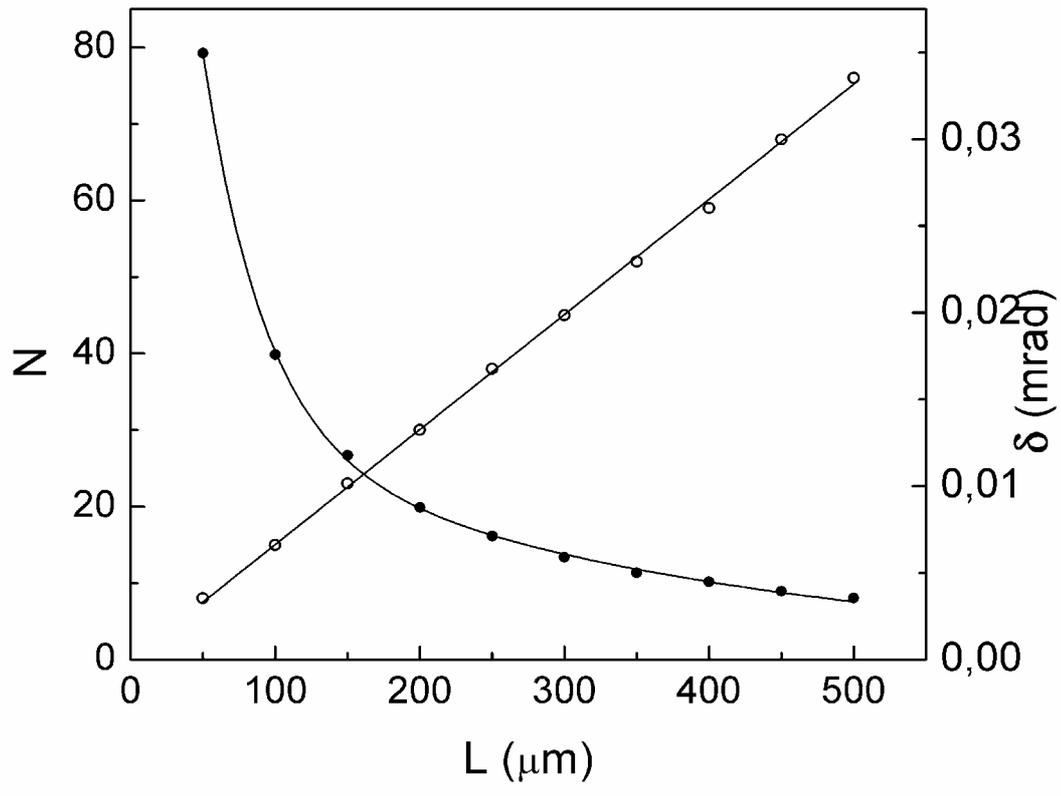

Figure 5